\documentclass[a4paper,12pt]{article}
\usepackage{graphicx}
\usepackage{default}

\title{Ignorance based inference of optimality \\in  thermodynamic processes}
\author{Preety Aneja and Ramandeep S. Johal \\
Indian Institute of Science Education and Research Mohali,\\
Knowledge City, Sector 81, Manauli P.O.,\\ Mohali-140306, Punjab, India.}
\newcommand{\be}{\begin{equation}}
\newcommand{\bea}{\begin{eqnarray}}
\newcommand{\bc}{\begin{center}}            
\newcommand{\ee}{\end{equation}}
\newcommand{\eea}{\end{eqnarray}}
\newcommand{\ec}{\end{center}}
\newcommand{\baa}{\begin{eqnarray*}}
\newcommand{\eaa}{\end{eqnarray*}}
\begin{document}
\baselineskip 18pt
\maketitle
\begin{abstract}
We derive ignorance based prior distribution to quantify incomplete
information and show its use to estimate the optimal 
work characteristics of a heat engine. 
\end{abstract}
Subjective probability or Bayesian inference methods seek to
quantify uncertainty due to incomplete prior knowledge about
the system \cite{Jeffreys1939, Jaynes2003}. 
The incomplete information is quantified
as a prior probability distribution, or simply known as a
prior and interpreted in the sense of degree of belief about
the likely values of the uncertain parameter. The choice
of an appropriate prior has been a crucial issue in
the Bayesian epistemology, which has hampered its development
and general acceptance for a long time. To outline the framework, 
consider a system which depends on two
parameters $T_1$ and $T_2$, where each parameter 
may lie within a specified range, $[T_-,T_+]$. Further, we are told 
that the values of $T_1$ and $T_2$ are constrained by a known relation,
 $T_1 = F(T_2)$,  so that given a value for one parameter, it 
implies a certain value of the other.   
Assume now that we have incomplete information about the system,
which refers to an ignorance about
the exact values of the above parameters.
In this paper, we present evidence that inference based on subjective ignorance,
has a close relation with optimal characteristics
of certain thermodynamic systems.
The present approach was recently proposed and applied \cite{Johal2010}
to quantify similar lack of information in models of heat engines \cite{GRD2, PRD1},
where the optimal expected behavior of the engine
was observed at certain well-known efficiencies, such
as Curzon-Ahlborn efficiency, $1-\sqrt{1-\eta_C}$, where
$\eta_C$ is the Carnot efficiency.

The essence of our approach is to assign prior probabilities for the likely
values of $T_1$ or $T_2 $. 
The assignment is guided by
the available prior information about the problem, such as 
the similar nature of these variables,
and  the fact that the problem is symmetric with respect to $T_1$ and $T_2$.
Thus for a certain pair of values, related by $T_1 = F(T_2)$,
one may only assign equal probabilities for these values,
in the range $[T_1, T_1 + dT_1]$ and $[T_2, T_2 + dT_2]$, respectively.
 This principle implies the following:
\be
P(T_1)dT_1 = P(T_2)dT_2.
\label{eqprob}
\ee
 Note that we have chosen
the same distribution function $P$ for both $T_1 $ and $T_2$.
The task then is to solve for the function $P$,
and find the estimate for $T_2$, defined as 
$\overline{T}_2 = \int T_2 P(T_2)dT_2$,
the average value over the prior. The corresponding 
estimate for $T_1$ is given by $T_1 = F(\overline{T}_2)$.
Any other quantity pertaining to this system
which is a function of $T_1$ and $T_2$, may be estimated
based on these values.

It is easy to visualise a physical analog of
the above scenario. 
Consider a pair of thermodynamic systems, 
identical in all aspects, except that
their initial temperatures are 
$T_+$ and $T_-$, respectively. 
Assume that $T_+ > T_-$. Let the fundamental 
thermodynamic relation of each system is given by
$S \propto U^{\omega_1}$, where the constant of proportionality
 may depend on some universal constants and/or volume, particle number
of the system. 
Using ${\partial S}/{\partial U}={1}/{T}$, we get:
$U \propto T^{{1}/{(1-\omega_1)}}$.
Alternately, we can write 
$S \propto T^{\omega_1/(1-\omega_1)}$.
We restrict to the case $0<\omega_1<1$, which implies
systems with a positive heat capacity.
 Some well-known
physical examples in this framework are the ideal Fermi gas ($\omega_1=1/2$),
 the degenerate Bose gas ($\omega_1=3/5$) and the black body radiation
($\omega_1=3/4$). Classical ideal gas can also
be treated as the limit, $\omega_1\to 0$.

Now assume that after some mutual interaction, the temperatures of 
the two systems become $T_1$ and $T_2$ and the relation between them is 
specified as follows:
\bea
T_1=({T_+}^{\omega} + {T_-}^{\omega}-{T_2}^{\omega})^{\frac{1}{\omega}},
\label{n11}
\eea
where $ \omega={\omega_1}/{(1-\omega_1)}$.
Physically, this dependence is obtained if we regard the interaction
as a reversible process, preserving the  thermodynamic entropy
of the composite system \cite{Callen1985}. This implies 
$\Delta S = \Delta S_1 + \Delta S_2 = 0$,
where $\Delta S \equiv S_{\rm fin}-S_{\rm ini}$. Moreover, one can 
extract work ($W$) in this process, which is  
 equal to the decrease in internal
energy ($U$) of the total system, $W = -\Delta U$, where $\Delta
U=U_{\rm fin}-U_{\rm ini}$. The expression for work (upto a
constant of proportionality) is:
\bea
W=({T_+}^{\frac{1}{1-\omega_1}} +
{T_-}^{\frac{1}{1-\omega_1}})-({T_1}^{\frac{1}{1-\omega_1}}
 + {T_2}^{\frac{1}{1-\omega_1}}).
\label{n}
\eea
Substituting the value of $T_1$ from Eq. (\ref{n11}) into Eq. (\ref{n}),
we may regard $W$ as a function of $T_2$ only:
\be
W={T_+}^{\frac{1}{1-\omega_1}}+{T_-}^{\frac{1}{1-\omega_1}}-({T_+}^\omega +
{T_-}^\omega-
{T_2}^\omega)^{\frac{1}{\omega_1}}-{T_2}^{\frac{1}{1-\omega_1}}.
\label{n1}
\ee
One may continue to extract more work till the two systems achieve a common
temperature $T_c$. We call this the optimal work extractable from 
the initial set up, where the final temperature of the systems are given by:
\be
T_c=\left(\frac{{T_-}^\omega+{T_+}^\omega}{2}\right)^{\frac{1}{\omega}}.
\label{j}
\ee
Now let us regard the incomplete information in the present context,
to arise from a lack of knowledge about the intermediate temperatures,
 $T_1$ and $T_2$. Specifying the physical process fixes only the  
functional relation between them, $T_1 = F(T_2)$.
As discussed in the Introduction, we 
invoke the device of two players A and B, who assign probabilities 
for $T_1$ and $T_2$ respectively, according to their degree of
belief. Then implementing the  criterion of Eq. (\ref{eqprob}),
we obtain the following normalised prior:
\be
P(T_2)=\frac{\omega {T_2}^{\omega-1}}{{T_+}^{\omega}(1-\theta^{\omega})},
\label{l1}
\ee
where $\theta={T_-}/{T_+}$. We have chosen to do the analysis
by using $T_2$ as the variable. The symmetry between $T_1$ and $T_2$,
in the work and entropy expressions, indicates that equivalently we can also
use the variable $T_1$.  

The expected value of $T_2$ is:
\be
\overline{T}_2=\int_{T_-}^{T_+}{T_2 P(T_2) dT_2 }.
\ee
In the following, we choose $T_+ =1$, for simplicity.
After solving the above integral, we obtain 
\be
\overline{T_2} = \omega_1\frac{(1-\theta^{\frac{1}{1-\omega_1}})}
{(1-\theta^{\frac{\omega_1}{1-\omega_1}})}.
\ee
The expected value of work can be found by substituting  $\overline{ T_2}$
in place of  $T_2$ in (\ref{n1}):
\be
{\overline W}_p=1-{\left[
(1+\theta^{\omega})
-\left(\omega_1
\frac{1-\theta^{{\frac{1}{1-\omega_1}}}}{1-\theta^{\frac{\omega_1}{1-\omega_1}}}
\right)
^{\omega}\right]}^{\frac{1}{\omega_1}}
+\theta^{\frac{1}{1-\omega_1}}-\left(\omega_1
\frac{1-\theta^{{\frac{1}{1-\omega_1}}}}{1-\theta^{\frac{\omega_1}{1-\omega_1}}}
\right)
^{\frac{1}{1-\omega_1}}.
\label{p1}
\ee
Here subscript $p$ refers to the power-law prior.
The input heat ($Q$) is given by the
difference of the initial and the final energies of the initially hotter system. 
We can estimate the value of this quantity similarly as above. The result is
\be
{\overline Q}_p
=1-\left[(1+\theta^{\omega})-
\left(\omega_1 \frac{1-\theta^{\frac{1}{1-\omega_1}}}
{1-\theta^{\frac{\omega_1}{1-\omega_1}}}\right)^{{\omega}}\right]^{\frac{1}{
\omega_1}}.
\label{q1}
\ee
Finally, the  efficiency is estimated by 
$\overline{\eta}_p = {{\overline W}_p}/ {\overline Q}_p$, and is
given by:
\be
\overline{\eta}_p=1+\frac{\theta^{\frac{1}{1-\omega_1}}-\left(\omega_1
\frac{1-\theta^{\frac{1}{1-\omega_1}}}
{1-\theta^{\frac{\omega_1}{1-\omega_1}}}\right)^{\frac{1}{1-\omega_1}}}{1-\left[
(1+\theta^{\omega})-
\left(\omega_1 \frac{1-\theta^{\frac{1}{1-\omega_1}}}
{1-\theta^{\frac{\omega_1}{1-\omega_1}}}\right)^{{\omega}}\right]^{\frac{1}{
\omega_1}}}.
\label{q11}
\ee
Now how does these estimated quantities compare with those arrived
from other priors? Or more significantly, how do the estimates
compare with some intrinsic features of the system, such as optimal 
work extracted?

For this purpose, we note the corresponding expressions for work and efficiency,
 when a uniform prior is used. This prior might appear as 
a natural choice in case the only information about the uncertain
parameter is its range, $[T_-,T_+]$. For $T_+=1$,
the uniform prior is given as 
\bea
P(T_2)dT_2&=&\frac{dT_2}{(1-\theta)}.
\eea
So the expected values of
different quantities are as follows:
\be
\overline{T_2}=\frac{1+\theta}{2},
\ee
\be
\overline{W}_u = 1-\left(1+\theta^{\omega}-\left(\frac{1+\theta}{2}
\right)^{\omega}\right)^{\frac{
1}{\omega_1}}+\theta^{\frac{1}{1-\omega_1}}-
\left(\frac{1+\theta}{2}\right)^{\frac{1}{1-\omega_1}},
\ee
\be
\overline{\eta}_u = 1+\frac{\theta^{\frac{1}{1-\omega_1}}-\left(\frac{1+\theta}{
2}\right)^{\frac{1}{1-\omega_1}}}
{1-\left(1+\theta^{\omega}-\left(\frac{1+\theta}{2}\right)^{\omega}\right)^{
\frac{1}{\omega_1}}}.
\ee
Here subscript $u$ refers to the uniform prior.

For optimal work, the final temperatures of the two systems are equal.
Substituting ${T_1} ={T_2}= T_c$
 in Eq. (\ref{n}):
\be
W_{o} = 1 + \theta^{\frac{1}{1-\omega_1}}-2
\left(\frac{1+\theta^{\frac{\omega_1}{1-\omega_1}}}{2}\right)^{\frac{1}{\omega_1
}}.
\label{q}
\ee
The efficiency at optimal work, is given by:
\be
\eta_{o}=1+\frac{\theta^{\frac{1}{1-\omega_1}}-(\frac{1+\theta^{\omega}}{2})^{
\frac{1}{\omega_1}}}
{1-(\frac{1+\theta^{\omega}}{2})^{\frac{1}{\omega_1}}}.
\ee
Fig. 1 shows the main result of the paper in the form of  plots 
comparing ${\overline W}_p$,  ${\overline W}_u$ and
$W_{o}$ for systems with different values of ${\omega_1}$.
\begin{figure}[h]
\begin{tabular}{ll}
\includegraphics[width=2in]{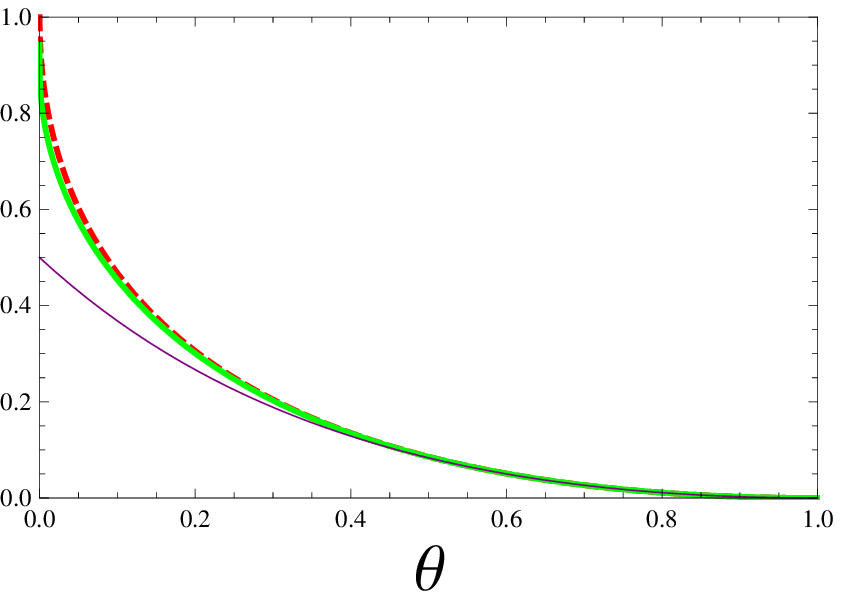} & \includegraphics[width=2in]{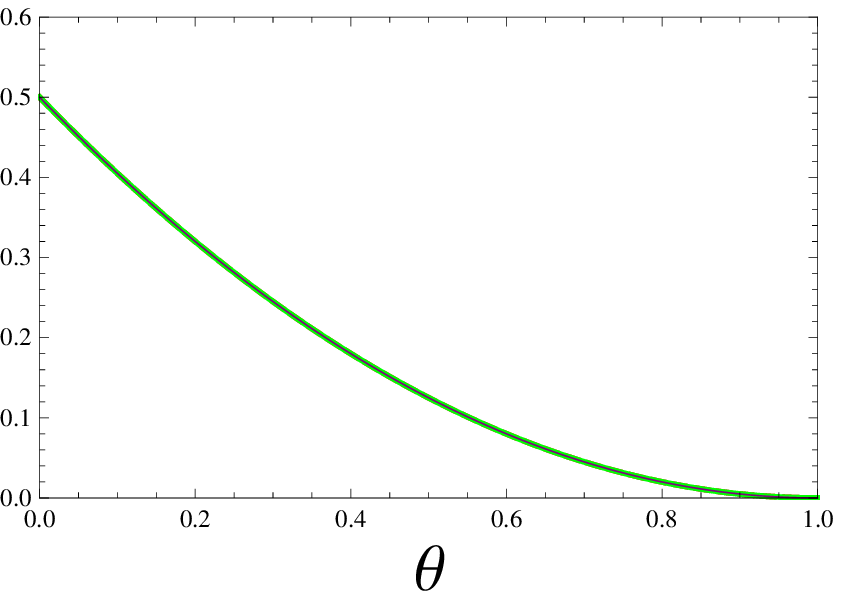}\\
\hspace{.9in}\small{\textit{(a)}} & \hspace{.95in}\small{\textit{(b)}} \\\\
\includegraphics[width=2in]{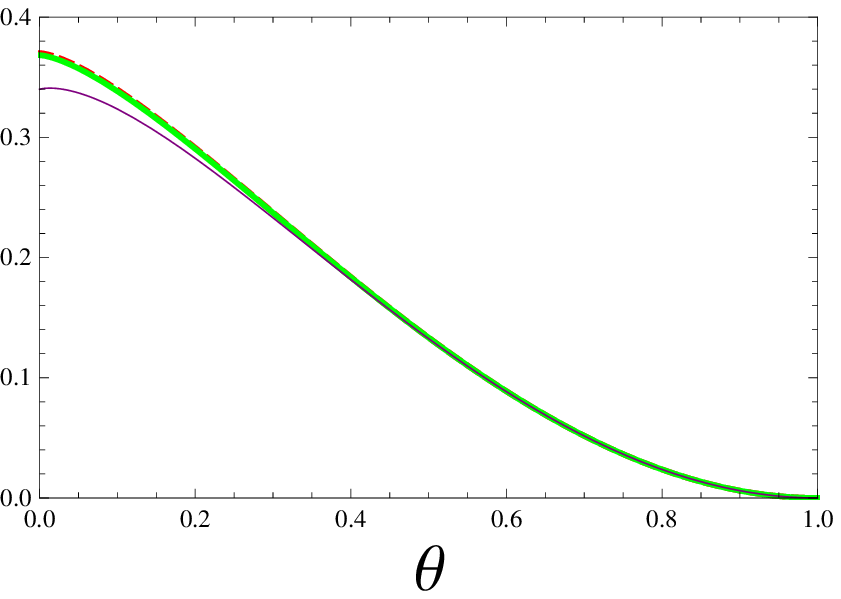} &\includegraphics[width=2in]{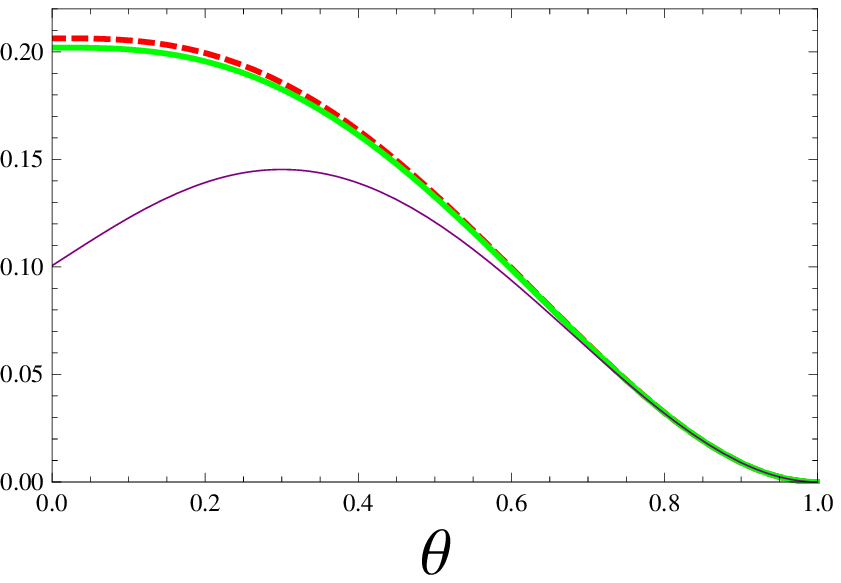}\\
\hspace{.95in}\small{\textit{(c)}} & \hspace{.95in}\small{\textit{(d)}}
\end{tabular}
\caption{Work as a function of $\theta$; (a) 
Ideal classical gas $(\omega_1\to0)$,
 (b) Ideal Fermi gas $(\omega_1={1}/{2})$,
(c) Degenerate Bose Gas $(\omega_1={3}/{5})$, 
(d) Black Body radiation $(\omega_1={3}/{4})$. The dashed curve is for $W_o$, 
thick curve is for $\overline{W}_p$, and thin curve is for 
 $ \overline{W}_u$.}
\end{figure}
We note in the near-equilibrium regime, there is a close proximity
 between the estimated work
(calculated with either prior, uniform or power-law) and
the optimal work.
In fact, expanding ${\overline W}_p$, ${\overline W}_u$ and $W_{o}$ about
$\theta=1$ up to third order, we obtain, in {\it each} case
\be
{\overline W}_p  =  \frac{1}{(1-\omega_1)}\frac{(1-\theta)^2}{4} 
+\frac{(1-2\omega_1)}{(1-\omega_1)^2}\frac{(1-\theta)^3}{8}+O[1-\theta]^4.
\ee
There is a remarkable agreement, upto third order, between 
the estimated and the optimal work. The estimate
appears to be insensitive to the choice of the prior, in this regime.
However, far from equilibrium, we  see in general, deviations
between the behavior predicted by different priors.  
Particularly, in the limit $\theta \rightarrow 0$,
\bea
{\overline W}_p & = & 1-\omega_1^{\frac{1}{1-\omega_1}}-
(1-\omega_1^{\frac{\omega_1}{1-\omega_1}})^{\frac{1}{\omega_1}},\\
{\overline W}_u & = & 1-\left(\frac{1}{2} \right )^{\frac{1}{1-\omega_1}}-
\left (1-\left( \frac{1}{2} \right)^{\frac{\omega_1}{1-\omega_1}}
\right)^{\frac{1}{\omega_1}}, \\
W_{o} & = & 1-\left( \frac{1}{2}\right)^{\frac{1-\omega_1}{\omega_1}}.
\eea
In general, we observe that the results from the power-law prior  are
very close to the optimal behavior. It is remarkable 
that an average over the ignorance probabilities 
can yield good estimates of the optimal characteristics
of the engine. 

We also compare the estimates for efficiency, with 
the efficiency at optimal work.
\begin{figure}[h]
\begin{tabular}{ll}
\includegraphics[width=2in]{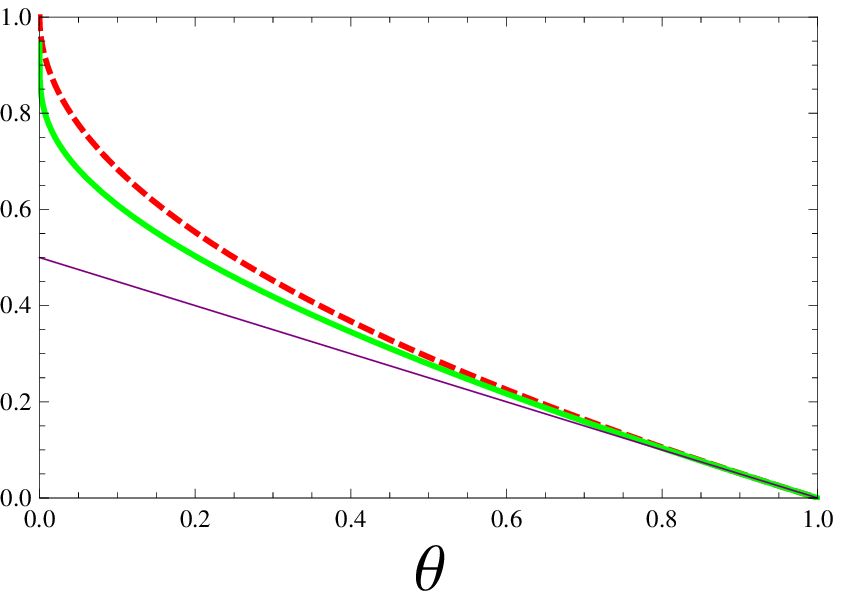} & \includegraphics[width=2in]{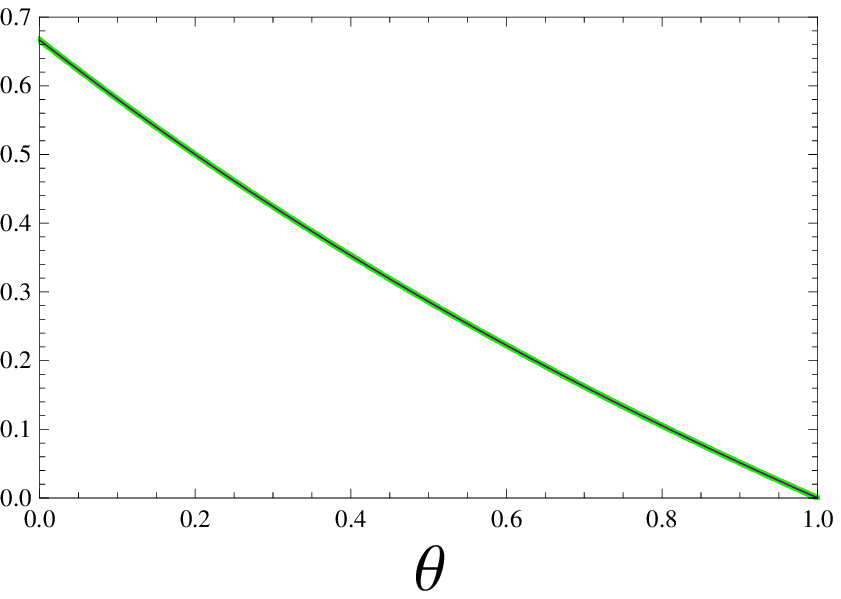}\\
\hspace{.9in}\small{\textit{(a)}} & \hspace{.95in}\small{\textit{(b)}} \\\\\\\
\includegraphics[width=2in]{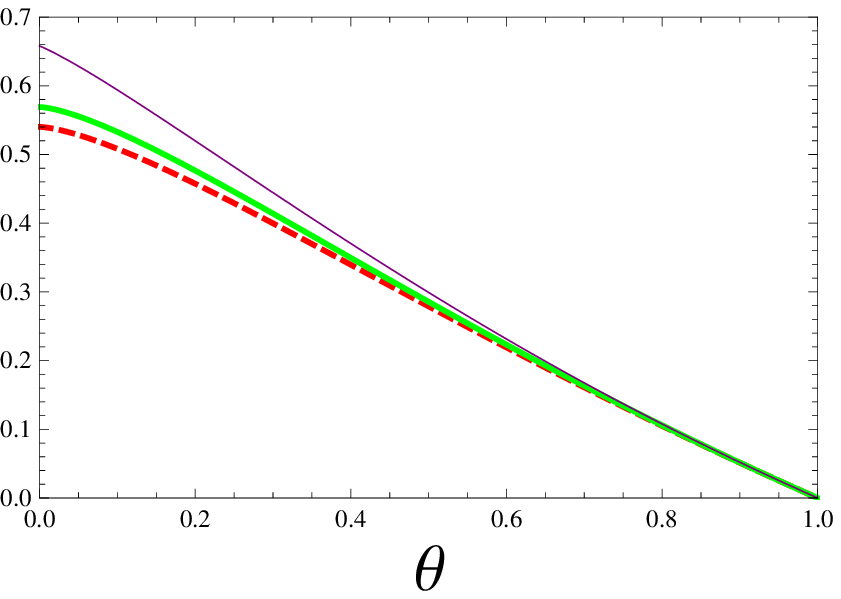} & \includegraphics[width=2in]{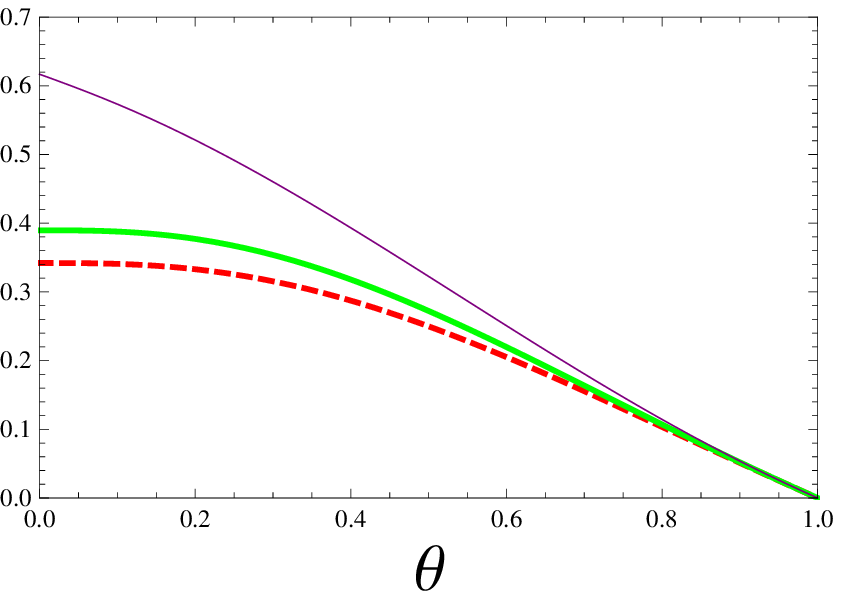}\\
\hspace{.95in}\small{\textit{(c)}} & \hspace{.95in}\small{\textit{(d)}}
\end{tabular}
\caption{Efficiency as a function of $\theta$; (a) Ideal classical gas,
(b) Ideal Fermi gas, (c) Degenerate Bose gas,
(d) Black Body radiation. 
Dashed curve is for $\eta_o$, thick 
curve is for $\overline{\eta}_p$ and the thin curve is for $\overline {\eta}_u$.}
\end{figure}
Expanding $\overline{\eta}_p$, $\overline{\eta}_u$ and $\eta_o$ about
$\theta=1$ up to 2nd order,
\bea
\overline{\eta}_p&=&\frac{\eta_{c}}{2}+\frac{(2-\omega_1)}{(1-\omega_1)}\frac{
\eta_{c}^2}{24 }+O[\eta_{c}]^{3}.\\
\overline{\eta}_u&=&\frac{\eta_{c}}{2}+\frac{\omega_1}{(1-\omega_1)}\frac{\eta_{
c}^2}{8}+O[\eta_{c}]^{3}.\\
\eta_{o}&=&\frac{\eta_{c}}{2}+\frac{\eta_{c}^2}{8}+O[\eta_{c}]^3.
\eea
It is also observed
that for $\omega_1 = 1/2$ (ideal Fermi gas), we have an exact
equality of both the work and the efficiency estimates  with 
the optimal characteristics (see Fig.1b, Fig2b). Actually, the power-law
prior becomes the uniform prior for this case. 
The efficiency at  optimal work
in this case gets simplified to $\eta_p = \eta_u = \eta_o = 2\eta_C/(4-\eta_C)$,
where $\eta_C = (1-\theta)$.
Interestingly, this expression also arises within the 
framework of stochastic thermodynamics applied to
a model of brownian heat engine \cite{Schmiedl2008}.

In the above comparison, we have seen that power-law prior with
a specific exponent, yields a better estimate for optimal work
extraction, as compared to
the uniform prior. A measure for the reliability of the estimates
 may be the variance of the estimated
parameter $T_2$, which  is evaluated as:
\be
V_p=\frac{\omega_1}{2-\omega_1}
\left(\frac{1-\theta^{\frac{2-\omega_1}{1-\omega_1}}}{1-\theta^
{\frac{\omega_1}{1-\omega_1}}}\right)-{\omega_1}^{2}\left(\frac{1-\theta^{\frac{
1}{1-\omega_1}}}
{1-\theta^{\frac{\omega_1}{1-\omega_1}}}\right)^{2},
\ee
and using the uniform prior, we have
\be
V_u = \frac{(1-\theta)^2}{12}.
\ee
Fig. 3 shows the comparison of the variance for both
the priors, for different systems.
\begin{figure}
\begin{tabular}{ll}
\includegraphics[width=2in]{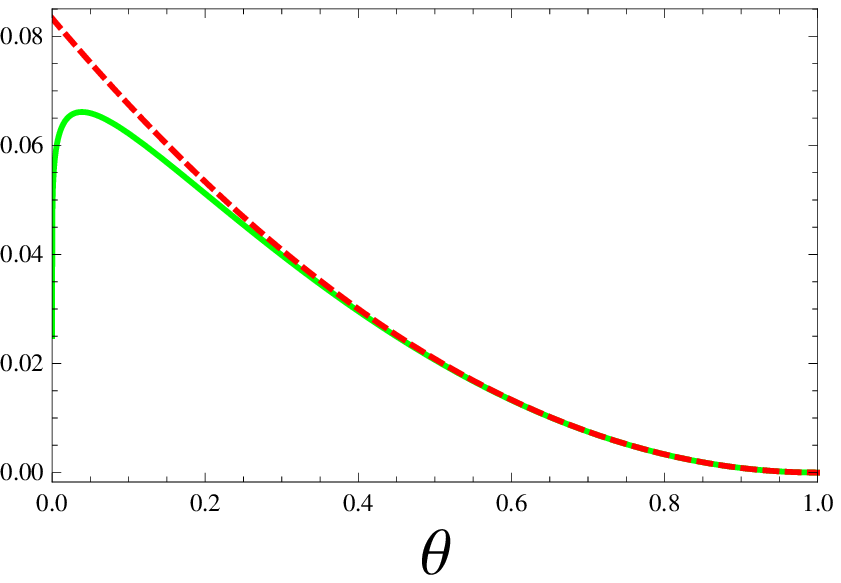} & \includegraphics[width=2in]{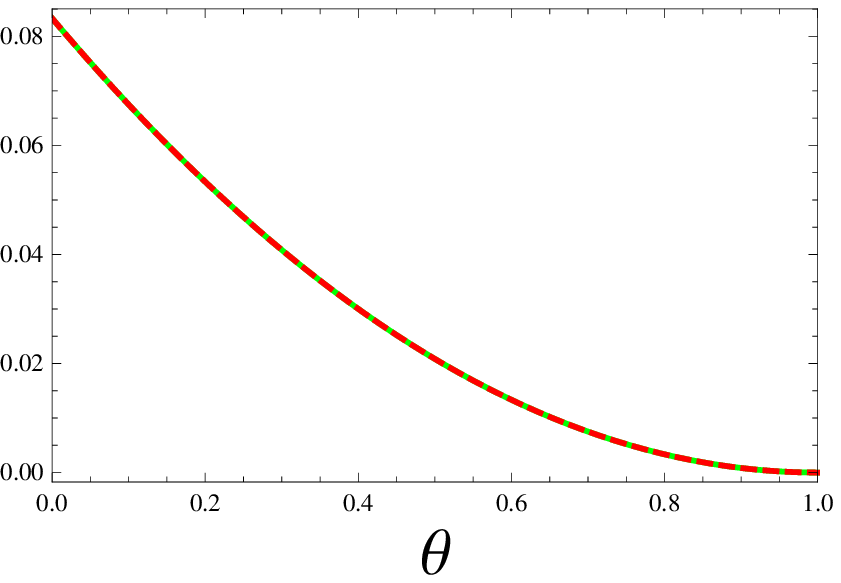}\\
\hspace{0.9in} \small{\textit{(a)}} & \hspace{0.95in}\small{\textit{(b)}} \\\\\\\
\includegraphics[width=2in]{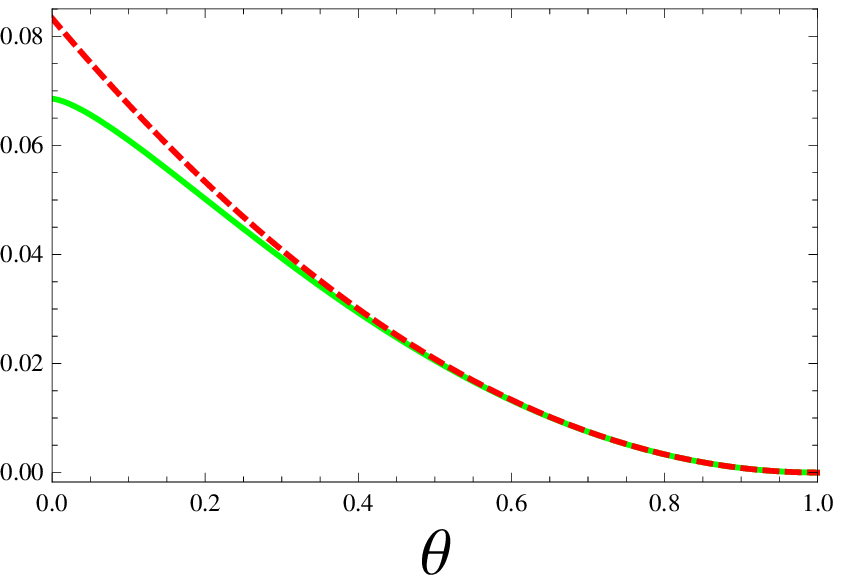} & \includegraphics[width=2in]{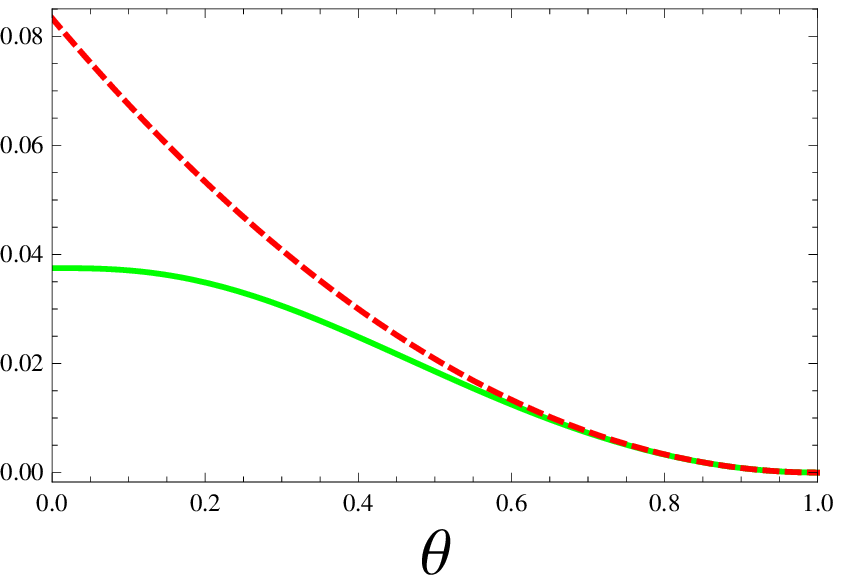}\\
\hspace{.95in}\small{\textit{(c)}}  & \hspace{.95in}\small{\textit{(d)}}
\end{tabular}
\caption{ Variance of $T_2$ as function of $\theta$. (a) Classical ideal gas
 (b) Fermi ideal gas  (c) Degenerate Bose gas, (d)
Black body radiation. Dashed curve is for uniform prior, while the solid 
curve is for power law prior.}
\end{figure}
In general, the variance of $T_2$ is less for power-law prior than for 
uniform prior, reflecting the higher reliability of the power-law prior estimates. 

Concluding, we find evidence suggesting the relevance
of ignorance based priors to infer the
optimal characteristics of a heat engine with two
finite reservoirs.
The final 
state in case of optimal work extraction is the equilibrium
state (minimum total energy) 
when  temperatures of both reservoirs become equal. The ignorance
based approach yields good estimates of the final state,
 from a consideration of the intermediate nonequilibrium 
states.
For small temperature differences, the results
are not sensitive to the prior and a uniform
prior is equally good to estimate the optimal behavior.
Far away from equilibrium, 
 the results are sensitive to the choice of a prior. So
 we have presented arguments for the choice of the prior.
 The characteristics of the engine
are estimated by averaging with the prior over all the 
values of its range.   The present approach might indicate
a subtle relation between the equilibrium and nonequilibrium states.
Can there be a possible relation with the recent fluctuation theorems,
 where the information on equilibrium
free energies is obtained by averaging over
nonequilibrium paths \cite{Jarzynski1997, Crooks1998}? 

R.S.J. acknowledges financial support from the 
Department of Science and Technology,
 India under the research project No. SR/S2/CMP-0047/2010(G).
P.A. acknowledges the grant of Junior Research Fellowship 
from University Grants Commission, India.

\end{document}